\documentclass[prl,floatfix,twocolumn,superscriptaddress,showpacs,amsfonts,amssymb]{revtex4}
\usepackage{graphicx}
\usepackage{dcolumn}
\usepackage{bm}
\begin{document}
\preprint{APS/123-QED}
\title{Competing periodicities in fractionally filled one-dimensional bands}

\author{P.C. Snijders}
\affiliation{Kavli Institute of NanoScience, Delft University of
Technology, 2628 CJ Delft, The Netherlands}
\author{S. Rogge}
\affiliation{Kavli Institute of NanoScience, Delft University of
Technology, 2628 CJ Delft, The Netherlands}
\author{H.H. Weitering}
\affiliation{Department of Physics and Astronomy, The University
of Tennessee, Knoxville, TN 37996, and Condensed Matter Sciences
Division, Oak Ridge National Laboratory, Oak Ridge, TN 37831, USA}
\date{\today}

\begin{abstract}
We present a variable temperature Scanning Tunneling Microscopy
and Spectroscopy (STM and STS) study of the Si(553)-Au atomic
chain reconstruction. This quasi one-dimensional (1D) system
undergoes at least two charge density wave (CDW) transitions at
low temperature, which can be attributed to electronic
instabilities in the fractionally-filled 1D bands of the
high-symmetry phase. Upon cooling, Si(553)-Au first undergoes a
single-band Peierls distortion, resulting in period doubling along
the imaged chains. This Peierls state is ultimately overcome by a
competing triple-period CDW, which in turn is accompanied by a
$\times2$ periodicity in between the chains. These locked-in
periodicities indicate small charge transfer between the nearly
half-filled and quarter-filled 1D bands. The presence and the
mobility of atomic scale dislocations in the $\times3$ CDW state
indicates the possibility of manipulating phase solitons carrying
a (spin,charge) of ($1/2,\pm e/3$) or ($0,\pm 2e/3$).

\end{abstract}
\maketitle
According to the Mermin-Wagner theorem \cite{Mermin66-68},
thermodynamic fluctuations preclude the formation of a long-range
ordered broken symmetry state in 1D, except at \textit{T} = 0 K
\cite{Rice73Pytte74}. For all practical purposes, however,
thermodynamic phase transitions may still be possible in finite
size 1D systems. Furthermore, fluctuations are inevitably
suppressed if the 1D chains are weakly coupled, or if the chains
are coupled to a substrate \cite{Rice73Pytte74,Lee04Shannon96}.
Prototypical 1D metallic systems like the transition metal
trichalcogenides, organic charge transfer salts, blue bronzes, and
probably all atomic Au-chain reconstructions on vicinal Si
substrates exhibit symmetry breaking phase transitions at finite
temperatures \cite{Gruner85,Yeom05}. For a band filling of $1/n$,
the phase transition opens up a gap in the single particle
excitation spectrum at wavevector $k_{F}=\pi/na$, and the
corresponding broken symmetry state adopts the new periodicity of
$\lambda=\pi/k_{F}=na$, where $a$ is the lattice parameter of the
high symmetry phase \cite{Gruner85}.

Fractional band fillings other than half filling provide an
interesting subset of 1D systems which often exhibit exotic
physical phenomena. Depending on the relative magnitude of
bandwidth and electron-electron interaction, CDW states often
compete with spin density waves, Mott insulating states, or a
Luttinger liquid state. Atomic-scale STM observations of surface phase transitions
 provide important insights into the complexity of symmetry
breaking phenomena in reduced dimensionality \cite{In}. For
instance, the recently reported $4\times1$-to-$8\times2$ phase
transition in quasi-1D indium chains on Si(111) \cite{In} involves
a gap opening in a complex triple band Peierls system, resulting
in a doubling of the periodicity along the atom chains. Another
recently discovered system with three fractionally-filled bands is
the Si(553)-Au surface. Angle-resolved photoemission spectroscopy
(ARPES) \cite{Crain0304} revealed three metallic bands but despite
theoretical efforts to understand the electronic structure
{\cite{Riikonen05,Crain0304} the atomic structure and real space
location of the surface state orbitals remains unknown.
Interestingly, the total band filling of this particular chain
system is $4/3$, \emph{i.e.} corresponding to $8/3$ electrons per
surface unit cell. Two bands have a filling of 0.56 and 0.51 each,
slightly more than half-filling. The third band has a filling of
0.27, in between one quarter and one-third filling.

In this Letter we present evidence for a defect mediated CDW
transition \cite{Hanno99} in Si(553)-Au accompanied by a metal
insulator transition. STM experiments reveal competing
periodicities as a function of temperature which can be mapped
onto the band structure of the high symmetry phase. STS data show
a gradual gap opening at low temperatures, which can be correlated
with successive gap openings in the three 1D bands. Interestingly,
phase slips are observed in the CDW condensate. These phase slips
should possess a fractional charge and a half integer or integer
spin. The chain-length can be tuned by manipulating the numerous
defects with the STM tip. This, in turn, suggests the feasibility
of studying and manipulating fractional charges at the atomic
scale.

The Si(553)-Au structure was prepared by
depositing 0.24 ML of Au at a rate of 0.005 ML/s with the
substrate held at 920 K, followed by thermal annealing at 1120 K
for 1 minute and slow cooling to room temperature or RT (1
K$s^{-1}$). STM and STS experiments were performed in an Omicron
variable temperature STM. All distances determined in STM images
were measured along the fast scan direction of the STM so as to
minimize possible effects of thermal drift.

Figure~\ref{fig:RT} shows an STM image of the surface taken at
room temperature (RT). Rather wide chains ($\sim0.8$~nm) with a
spacing of 1.48 nm are observed. The chains are cut by point
defects appearing as vacancies in both filled state and empty
state images. In the empty state image, these vacancies appear
larger than in the filled state image, confirming the observations
reported in Ref.~\onlinecite{Crain05}. Note, however, that fewer
defects are present as compared to other work
\cite{Crain0304,Crain05}, indicating that the defect concentration
can possibly be varied by controlling the annealing history of the
surface despite propositions that the defects might be intrinsic
to the surface as stabilizing charge dopants~\cite{Crain0304}. The
insets show high resolution dual bias images. Clearly, the chains
are composed of two rows of atoms, showing a \emph{zig-zag}
structure in the empty state and a \emph{ladder} configuration in
the filled state. In the empty state image all atom spacings are
nearly equal to the bulk spacing of Si, $a\sim3.8$~\AA~ resulting
in bond angles near 60 degrees. In the filled state image the
spacing along the chain is equal to $a$ whereas the spacing
\emph{perpendicular} to the chains is $\sim4.7$~\AA. These data
seem to be inconsistent with structure models that place Si
honeycomb chains \cite{Hanno98} near the step edges
\cite{Crain0304,Riikonen05} and with the single atom row structure
suggested in Ref.~\onlinecite{Crain05}.

\begin{figure}[ht]
  \centering
  \includegraphics[width=\columnwidth]{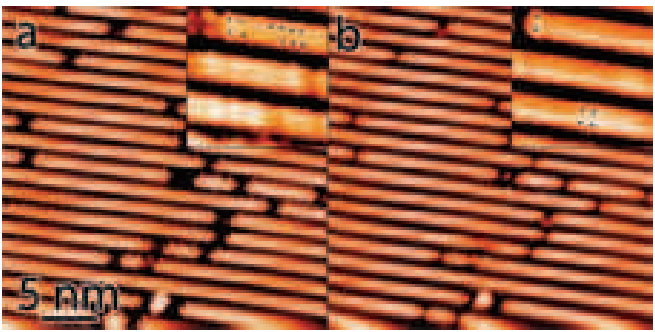}
  \caption{(color online) (\emph{a}) Empty state and (\emph{b}) filled state ($\pm 0.5
V, 50 pA$) STM images taken simultaneously at RT. Insets show
magnifications. The structure in the chains is indicated with
dots.} \label{fig:RT}
\end{figure}

Upon cooling to $\sim40$~K, the zig-zag features are no longer
observed. Instead, an up-down buckling with a ladder structure in
a \emph{tripled} unit cell is observed (Fig.~\ref{fig:40K}),
resulting in two features of about $1.5a$ length and widths
comparable to those observed at RT. The corrugation of the filled
state image is in anti-phase with that of the empty state image;
at the location of the intensity maxima in the empty state image,
a small dip exists in the filled state image. This indicates that
the system has condensed into a CDW with tripled periodicity,
commensurate with the substrate lattice. Additionally, the
features in the valleys between the chains visible in the empty
state are completely ordered with a \emph{doubled} period of $2a$,
resulting in a unit cell of $6\times1$.

The $I-V$-curves and their numerical derivatives are shown in
Fig.~\ref{fig:STS}. The $I-V$-curve at RT exhibits a significant
slope at zero bias, confirming the metallicity observed in ARPES
experiments \cite{Crain0304}. In contrast, at 40 K the $I-V$-curve
is flat at zero bias indicating semiconducting behavior. We
determine the size of the gap from the derivative of the
$I-V$-curves as displayed in the inset of Fig.~\ref{fig:STS}; we
infer that the excitation gap is symmetric and $\sim150$~meV wide.
Thus, the condensation of this CDW is accompanied by a
metal-insulator transition (MIT). Although CDW instabilities in
related surface systems are often manifested by an order-disorder
transition~\cite{Avila99}, the present observations are most
straightforwardly interpreted in terms of a displacive CDW
transition.

\begin{figure}[ht]
  \centering
  \includegraphics[width=6cm]{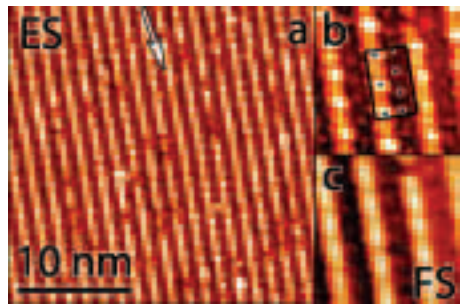}
  \caption{(color online) (\emph{a}) Empty state (1 V, 100 pA) STM image at 40
  K and magnification (\emph{b}). In (\emph{b}) a $6\times1$ unit cell is
indicated with
  combined $\times3$ and $\times2$ features. (\emph{c}) Filled state (-1 V,
  100 pA) STM image corresponding to (\emph{b}).}
  \label{fig:40K}
\end{figure}

\begin{figure}[ht]
  \centering
  \includegraphics[width=6cm]{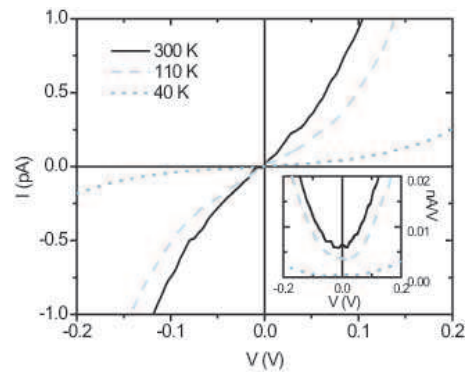}
  \caption{Area averaged STS curves measured at indicated temperatures. Inset: numerical
derivatives of the STS curves.}
  \label{fig:STS}
\end{figure}

The condensation scenario involving all three 1D bands is
elucidated by STM experiments at temperatures intermediate between
RT and 40 K. Fig.~\ref{fig:70-110K} shows empty state images
measured at 70 K and at 110 K. At 70 K a vague tripled corrugation
is visible in nearly all of the chains, with significantly
enhanced intensity near defects. In the middle of longer chains
segments the bulk period of $3.8$~\AA~is still visible through the
superimposed (vague) tripled periodicity. At 110 K, we observe a
\emph{doubled} periodicity in the bulk of most chains, but, again
near defects a tripled periodicity decaying into the chains is
observed. From these data at higher temperatures, it is evident
that the CDW present at 40 K nucleates from the defects and
spreads along the chains with decreasing temperature. Even at RT,
it is still possible to discern charge density oscillations
emanating from the defects. These oscillations have been
attributed to zero-dimensional end state effects \cite{Crain05}.
Alternatively, the change in apparent height of the chain adjacent
to a defect, \emph{i.e.} a depression over a distance of $\sim 1.5
a$ next to the defect followed by a brighter segment (see
Fig.~\ref{fig:RT}(\emph{a}):inset), and the non-metallic character
of the chain ends established in Ref. \cite{Crain05}, are fully
consistent with the interpretation of a CDW precursor, similar to
that observed in Sn/Ge(111) \cite{Hanno99}.

\begin{figure}[ht]
  \centering
  \includegraphics[width=6cm]{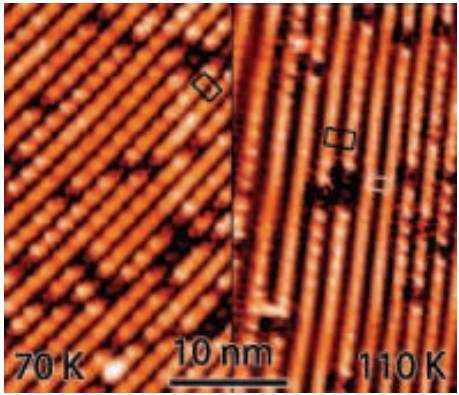}
  \caption{(color online) Empty state (1 V, 50 pA) STM images at 70 K
and 110 K. $\times3$ and $\times2$ unit cells indicated in black
and white, respectively.} \label{fig:70-110K}
\end{figure}

As mentioned above, the band structure of this surface measured by
ARPES contains three metallic bands \cite{Crain0304}; two bands
have a Fermi wave vector near half-filling, the third band
contains $0.27 \times 2 = 0.54$ electrons adding up to a total
filling of $4/3$. None of the bands crosses the Fermi energy
exactly at a wave vector $2\pi/(n\times a)$. Nevertheless, we
observe three commensurate periodicities evolving as a function of
temperature: at 110 K a clear $\times2$ period is detected
\emph{in} the chains, at 40 K a $\times2$ period is observed
\emph{between} the chains accompanied by a $\times3$ period in the
chains. Tentatively, we assign the three observed periods to
electronic instabilities in the three metallic bands of the high
symmetry phase, which would locate the orbitals of the 0.51 and
0.27 filled bands on top of the zig-zag chains and the orbitals of
the 0.56 filled band in between the chains. At 110 K, the doubled
periodicity along the chains originates from a CDW transition in
the 0.51 filled band; the other two bands remain metallic. STS
measurements support this scenario, still showing metallic
behaviour at 110 K but with reduced slope at zero bias, indicating
a reduced DOS at the Fermi energy, see Fig.~\ref{fig:STS}. ARPES
at temperatures $<100$ K show weak backfolding of the 0.27 filled
band at the $\times2$ zone boundary \cite{Crain0304}, also
indicating a doubled periodicity in this band. The $\times3$ CDW
at lower temperatures can then be rationalized by a small charge
transfer of 0.06 electron from the 0.56 filled band to the 0.27
filled band, filling the latter to $1/3$
\endnote{According to Ref. \cite{Schrieffer73}, a commensurate CDW
in a system with incommensurate Fermi wave-vectors can be obtained
by locking the wavelength of the CDW to the lattice so as to be
commensurate. However, then it would be expected that the 0.27
filled band would lock to a $\times4$ period corresponding to a
filling of 0.25.}. The possible CDW precursor near the chain ends
at RT can also be explained by this charge transfer: at RT a
$\times2$ period exists locally in between the chains near defects
that are located on the chains \cite{Crain05}. This strongly
suggests that already at RT the 0.56 filled band, located in the
valleys between the chains, is doping the chains near the defects
to $1/3$ filling. With this charge transfer the \emph{total} band
filling of the surface then remains constant at $4/3$. Note that
this evolution in band structure might also explain the
transformation of the features in the empty state STM image; from
a zig-zag structure at RT to a ladder structure at 110 K and
below.

Surprisingly, the competition between periodicities \emph{inside}
the chains is eventually won by a $\times3$ CDW at the lowest
temperature studied. Longer wavelength periods are progressively
more difficult to fit into chains with randomly placed fixed
defects. Furthermore, a $1/3$ filled parabolic band has a higher
DOS at the Fermi energy as compared to a $1/2$ filled band, which
according to a simple BCS theory argument should result in a
higher $T_{c}$ for the $1/3$ filled band. The magnitude of the
interchain coupling for the different bands, though being fairly
low as compared to other (bulk) 1D compounds \cite{Crain0304},
might provide insight into this seeming contradiction. It is well
known that finite interchain coupling reduces the transition
temperature and indeed ARPES measurements indicate a five times
larger interchain coupling for the $1/3$ filled band than for both
of the $1/2$ filled bands. This could explain why the $\times3$
CDW sets in at a lower temperature than the period doubling CDW
observed at 110 K \endnote{An alternative explanation is that the
relatively large charge transfer into the 0.27 filled band induces
significant more lattice strain, thereby lowering its T$_{c}$ as
compared to the $\times2$ CDW.}.

Finite interchain coupling should be discernible in the STM images
through definite phase relations between periodicities in adjacent
chains, provided that the temperature is low enough. Careful
analysis of the  empty state STM images at 40 K indeed reveals
small domains of up to three or four chains width, showing a
constant phase relation, but no long range order is detected.
However, as can be observed in Fig.~\ref{fig:40K}(\emph{a}, arrow)
even at 40 K there are chains which do not exhibit a fully
developed CDW; the triple periodicity is not long range ordered,
despite the fact that a fully developed triple period would fit
the length of this chain segment. This indicates that the system
is affected by a significant amount of interchain coupling. This
is further illustrated in Fig.~\ref{fig:jump}(\emph{a}). Two
fairly long chains ($>25$ nm) are visible. Starting from the top
both chains lack a clear corrugation, but both chains develop a
definite up-down corrugation towards the bottom. The white bars
illustrate a phase slip of $2\pi/3$ in the right chain,
immediately followed by two similar phase slips in the left chain.
These phase slips can not be explained by a $\times3$ CDW mismatch
in a finite chain segment because both chains show a region of
small corrugation at the top (and beyond, not shown) with an
ill-defined phase so the CDW is not phase locked by defects.
Therefore the origin must be related to interchain coupling;
apparently, a phase slip in one chain can induce phase slips in
neighboring chains via interchain coupling. Interestingly, a phase
slip (phase soliton) in a $1/3$ filled band CDW carries a
(spin,charge) of ($1/2,\pm e/3$) or ($0,\pm 2e/3$)
\cite{Su81Kuhn95}.

Fractionally charged phase slips have been studied mostly
theoretically. Only in polyacetylene, a phase slip with a
fractional charge of $e/2$ has been observed by conductivity and
NMR experiments \cite{Su80}. To the best of our knowledge
fractionally charged phase slips in CDWs away from half filling
have never been observed directly by imaging. Their presence opens
up the possibility to study and manipulate fractional charges at
the atomic scale. They can possibly be manipulated by deliberately
creating defects in the chains so as to generate a misfit for the
$\times3$ CDW. In our STM experiments at 40 K defects sometimes
jump to different locations during imaging. For example,
Figs.~\ref{fig:jump}(\emph{b}) and (\emph{c}) show two sequential
($\Delta t=180s$) filled state images from the same area. The
white circle and the arrow show a group of three defects which has
moved two chains to the left. This behavior suggests that these
defects consist of Si adatoms~\cite{Crain0304,Riikonen05}, which
have been relocated by tip induced migration.

\begin{figure}[ht]
  \centering
  \includegraphics[width=6cm]{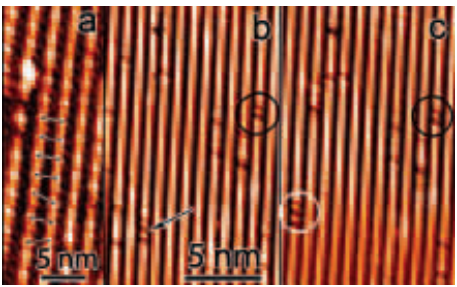}
  \caption{(color online) STM images at 40 K. (\emph{a}) Empty state (1 V, 100 pA) image
  showing phase slips in the $\times3$ CDW. (\emph{b}) and
  (\emph{c}): filled state (-1 V, 50 pA) images showing
  relocation of defects. Black circles: markers. Arrow: old defect
  position. White circle: new defect position.} \label{fig:jump}
\end{figure}

In conclusion, we have presented an STM and STS study of a chain
structure with fractional band fillings. Competing periodicities
are observed as a function of temperature resulting in a defect
mediated CDW at 40 K. The results can be mapped onto the band
structure of the high symmetry phase. The presence and mobility of
the chain dislocations in the CDW state indicates the possibility
of studying and possibly manipulating fractionally charged
solitons with an STM tip. The availability of other vicinal Si-Au
chain structures with tunable interchain coupling \cite{Crain0304}
would provide a promising playground for 1D physics, accessible in
real space. Note added: After the submission of our manuscript, we
became aware of the paper by Ahn et al.~\cite{Ahn05}. Their
observations are consistent with ours.

We thank Prof. Franz Himpsel for providing the Si(553) wafer. This
work is sponsored in part by NSF under contract No. DMR-0244570,
the Stichting voor Fundamenteel Onderzoek der Materie and the
Royal Netherlands Academy of Arts and Sciences. We thank T.M.
Klapwijk for his stimulating support. Oak Ridge National
Laboratory is managed by UT-Battelle, LLC, for the US Department
of Energy under contract No. DE-AC-05-00OR22725.


\begin{thebibliography}{99}
\bibitem{Mermin66-68}
N.D. Mermin, H. Wagner, Phys. Rev. Lett. {\bf 17}, 1133 (1966),
N.D. Mermin, Phys. Rev. {\bf 176}, 250 (1968).
\bibitem{Rice73Pytte74}
M.J. Rice, S. Str\"{a}ssler, Solid State Commun. {\bf 13}, 1389
(1973), E. Pytte, Phys. Rev. B {\bf 10}, 2039 (1974).
\bibitem{Lee04Shannon96}
M. Lee, E.A. Kim, J.S. Lim, M.Y. Choi, Phys. Rev. B {\bf 69},
115117 (2004), N. Shannon, R. Joynt, J. Phys.: Condens. Matter
{\bf 8}, 10493 (1996).
\bibitem{Gruner85}
G. Gr\"{u}ner, A. Zettl, Physics Reports {\bf 119}, 117 (1985).
\bibitem{Yeom05}
H.W. Yeom \emph{et al.}, Phys. Rev. B {\bf 72}, 035323 (2005).
\bibitem{In}
H.W. Yeom, K. Horikoshi, H.M. Zhang, K. Ono, R.I.G. Uhrberg, Phys.
Rev. B {\bf 65}, 241307(R) (2002), S.J. Park, H.W. Yeom, S.H. Min,
D.H. Park, I.W. Lyo, Phys. Rev. Lett. {\bf 93}, 106402 (2004), G.
Lee, J. Guo, E.W. Plummer, Phys. Rev. Lett. {\bf 95}, 116103
(2005), H. Morikawa, I. Matsuda, S. Hasegawa, Phys. Rev. B {\bf
70}, 085412 (2004), J. Guo, G. Lee, E.W. Plummer, Phys Rev. Lett.
{\bf 95}, 046102 (2005).
\bibitem{Crain0304}
J.N. Crain \emph{et al.}, Phys. Rev. Lett. {\bf 90}, 176805
(2003), J.N. Crain \emph{et al.}, Phys. Rev. B. {\bf 69}, 125401
(2004).
\bibitem{Riikonen05}
S. Riikonen, D. Sanchez-Portal, Nanotechnology {\bf 16}, S218
(2005).
\bibitem{Hanno99}
H.H. Weitering \emph{et al.}, Science {\bf 285}, 2107 (1999).
\bibitem{Crain05}
J.N. Crain, D.T. Pierce, Science {\bf 307}, 703 (2005).
\bibitem{Hanno98}
S.C. Erwin, H.H. Weitering, Phys. Rev. Lett. {\bf 81}, 2296
(1998).
\bibitem{Avila99}
J. Avila \emph{et al.}, Phys. Rev. Lett. {\bf 82}, 442 (1999).
\bibitem{Schrieffer73}
J.R. Schrieffer, Nobel Symposium 24 (Academic, New York and
London, 1973), p. 142.
\bibitem{Su81Kuhn95}
W.P. Su, J.R. Schrieffer, Phys. Rev. Lett. {\bf 46}, 738 (1981),
C. Kuhn, J. Phys. Condens. Matt. {\bf 7}, 6221 (1995).
\bibitem{Su80}
W.P. Su, J.R. Schrieffer, A.J Heeger, Phys. Rev. B {\bf 22}, 2099
(1980).
\bibitem{Ahn05}
J.R. Ahn \emph{et al.}, Phys. Rev. Lett. {\bf 95}, 196402 (2005)
\end{thebibliography}

\end{document}